# Effect of Ge-doping on the short-wave, mid- and far-infrared intersubband transitions in GaN/AlGaN heterostructures

**Caroline B Lim[1], Akhil Ajay[1], Jonas Lähnemann[2], Catherine Bougerol[3] and Eva Monroy[1]**

[1] University Grenoble-Alpes, CEA, INAC, PHELIQS, 17 av. des Martyrs, 38000 Grenoble, France

[2] Paul-Drude-Institut für Festkörperelektronik, Hausvogteiplatz 5-7, 10117 Berlin, Germany

[3] University Grenoble-Alpes, CNRS, Institut Néel, 25 av. des Martyrs, 38000 Grenoble, France

## Abstract

This paper assesses the effects of Ge-doping on the structural and optical (band-to-band and intersubband) properties of GaN/AlGaN multi-quantum wells designed to display intersubband absorption in the short-wave, mid- and far-infrared ranges (SWIR, MIR, and FIR, respectively). The standard *c*-plane crystallographic orientation is considered for wells absorbing in the SWIR and MIR spectral regions, whereas the FIR structures are grown along the nonpolar *m*-axis. In all cases, we compare the characteristics of Ge-doped and Si-doped samples with the same design and various doping levels. The use of Ge appears to improve the mosaicity of the highly lattice-mismatched GaN/AlN heterostructures. However, when reducing the lattice mismatch, the mosaicity is rather determined by the substrate and does not show any dependence on the dopant nature or concentration. From the optical point of view, by increasing the dopant density, we observe a blueshift of the photoluminescence in polar samples due to the screening of the internal electric field by free carriers. In the intersubband absorption, on the other hand, there is a systematic improvement of the linewidth when using Ge as a dopant for high doping levels, whatever the spectral region under consideration (i.e. different quantum well size, barrier composition and crystallographic orientation).

## 1. Introduction

GaN/AlGaN nanostructures have shown their potential for the development of new intersubband (ISB) optoelectronic devices, with the possibility to cover almost the whole infrared (IR) spectrum [1,2]. Their large conduction band offsets and sub-picosecond ISB relaxation times support the development of ultrafast photonic devices operating at telecommunication wavelengths [3–5]. Additionally, the large energy of the longitudinal-optical phonon in GaN (92 meV, 13 μm) opens prospects for THz quantum cascade lasers operating at room-temperature [6,7]. GaN is transparent for most of the IR spectral region, namely for wavelengths longer than 360 nm, except for the Reststrahlen band (phonon absorption band) located between 9.6 and 19 μm. Using $c$-plane GaN/AlN quantum wells (QWs), the ISB absorption can be tuned in the 1.0−3.5 μm wavelength range by changing the QW thickness from 1 nm to 7 nm [8–12]. To shift the absorption towards longer wavelengths, it is necessary to reduce the polarization-induced internal electric field in the QWs, which can be attained by using ternary AlGaN barriers with reduced Al mole fraction. Varying the geometry and composition of the barriers, the ISB absorption can be tailored to cover the range up to 10 μm [13–18]. Reducing the ISB transition energy below 60 meV (wavelength > 20 μm) requires further band engineering to compensate the internal electric field in the QWs, which is only achieved by implementation of complex multi-layer QW designs [19–22]. The use of nonpolar crystallographic orientations, particularly the $m$-plane, is a promising alternative to obtain GaN/AlGaN QWs without internal electric field [23–25].

The development of ISB devices requires a fine tuning of the n-type doping density, particularly in photodetectors and switches, where the first quantum-confined energy level of the conduction band has to be populated to allow electron transitions to higher levels. To measure ISB transitions in GaN/AlN QWs, large n-type doping densities ($N_D > 1\times10^{18}$ cm$^{-3}$) are required. The charge density has a significant impact on the transition energy and linewidth due to many-body effects such as depolarization shift and exchange interaction, as well as scattering by ionized impurities [8,23,26–29]. Furthermore, such doping concentrations can introduce strain, eventually leading to structural defects [30] or roughness at the heterointerfaces [31], which constitutes an additional source of carrier scattering. In general, studies of ISB transitions in III-nitrides have been performed using Si as n-type dopant, even though it is known that it introduces tensile strain which can lead to a structural degradation [30]. Ge is considered as an alternative n-type dopant, with improved performance in comparison to Si for doping levels in excess of $10^{19}$ cm$^{-3}$ [32–35]. Thanks to its ionic radius and metal-nitrogen bond length similar to that of Ga, the Ge occupancy of a Ga lattice site is expected to induce far less lattice distortions compared with Si.

In this paper, we assess the effect of Ge doping on the structural quality, band-to-band and ISB properties of polar *c*-plane GaN/AlGaN multi-quantum-wells (MQWs) designed for ISB absorption in the short-wave (SWIR, 1.4-3 μm) and mid-infrared (MIR, 3-5 μm) ranges, and of nonpolar *m*-plane GaN/AlGaN MQWs, designed for ISB absorption in the far-infrared (FIR, >20 μm) domain. We systematically compare identical structures doped with Ge and Si.

## 2. Methods

The band structure and quantum-confined electron levels were modeled using the Nextnano3 8×8 k.p self-consistent Schrödinger-Poisson solver [36] with the material parameters described in ref. [37]. The structure was treated in one dimension. The simulation covered a section of the MQW superlattice including three QWs, with periodic boundary conditions. With this simulation cell, we can confirm that the periodicity of the structure is correctly treated by checking that the results are identical in the three QWs.

The MQW structures were synthesized by plasma-assisted molecular-beam epitaxy (PAMBE). Thanks to its low growth temperature, this technique is particularly adequate to the fabrication of heterostructures with chemically abrupt interfaces [38]. Furthermore, it offers compatible growth conditions for the *c*-plane and the *m*-plane of GaN [39], and Si and Ge can be incorporated as dopants during the growth process, without perturbation of the growth kinetics [35,40]. For *c*-plane samples, the substrates were either 1-µm-thick AlN-on-sapphire templates with a dislocation density $\approx 10^9$ cm$^{-2}$ (GaN/AlN MQW structures) or 4-µm-thick GaN-on-Si(111) templates with a dislocation density $<5 \times 10^9$ cm$^{-2}$ (GaN/AlGaN MQW structures), both deposited by metalorganic vapor phase epitaxy. For *m*-plane samples, the substrates were 5 mm × 20 mm semi-insulating *m*-GaN platelets sliced from (0001)-oriented GaN boules synthesized by hydride vapor phase epitaxy (resistivity $>10^6$ Ωcm, dislocation density $<5 \times 10^6$ cm$^{-2}$). In all cases, the growth was performed under slightly Ga-rich conditions [24,25,39,41], at a substrate temperature of 720°C, and with a nitrogen-limited growth rate of 0.5 ML/s ($\approx$ 450 nm/h). A 200-nm-thick, non-intentionally-doped GaN buffer layer was deposited prior to the MQW, which was capped with 30-50 nm of AlGaN using the same Al content as for the barriers. The GaN wells were doped with either Si or Ge, at concentrations in the $5\times10^{11}$ cm$^{-2}$ to $6\times10^{13}$ cm$^{-2}$ range, as indicated in the list of samples in Table 1. The doping level was calibrated by Hall effect measurements in reference *c*-plane GaN:Si and GaN:Ge layers grown on AlN-on-sapphire templates.

The surface morphology of the layers was studied by atomic force microscopy (AFM) in the tapping mode using a Dimension 3100 system. The periodicity and structural properties of the MQWs were measured by high-resolution x-ray diffraction (HR-XRD) using a Rigaku SmartLab x-ray diffractometer with a 4-bounce Ge(220) monochromator and a 0.114 degree collimator in front of the detector. Samples were analyzed by high-angle annular dark-field scanning transmission electron microscopy (HAADF-STEM) performed in an FEI Titan Themis microscope operated at 200 kV. Cathodoluminescence (CL) measurements at 10 K were performed in a Zeiss Ultra-55 field-emission scanning electron microscope fitted with a Gatan MonoCL4 system, using an acceleration voltage of 3 kV and a current of about 400 pA. The emission from the sample was collected by a parabolic mirror and guided into a grating monochromator equipped with a liquid-nitrogen-cooled charge-coupled device (CCD) camera. Spectral line-scans were recorded with an integration time of 1 s per spectrum. Photoluminescence (PL) spectra were obtained by excitation with a continuous-wave solid-state laser ($\lambda$ = 244 nm), with an excitation power around 100 µW focused on a spot with a diameter around 100 µm. The emission from the sample was collected by a Jobin Yvon HR460 monochromator equipped with an ultraviolet-enhanced CCD camera. To assess the ISB properties of the MQWs, transmission measurements employing Fourier transform infrared (FTIR) spectroscopy were performed using a Bruker V70v spectrometer. For characterization in the SWIR and MIR ranges, the FTIR spectrometer was equipped with a halogen lamp and a nitrogen-cooled mercury–cadmium–telluride detector. All samples were polished at 45° (sapphire and *m*-GaN substrates) or at 30° (Si substrates) to form multipass waveguides allowing 4–5 interactions with the active region. For characterization in the FIR, we used a mercury lamp, a Si beam splitter and a helium-cooled Si bolometer. In the latter case, two pieces of each sample were placed face-to-face on the cold finger of a helium-cooled cryostat. All the samples were tested in transmission mode using a polarizer adapted to the targeted infrared region to discern between the transverse-electric (TE) and transverse-magnetic (TM) polarized light.

## 3. Experimental results

### A. SWIR absorption in *c*-plane GaN/AlN MQWs

We have studied a series of structures containing 25 periods of GaN/AlN (1.8 nm/3 nm) MQWs designed to show ISB absorption at 0.729 eV (1.70 µm) for low doping levels (samples S11 to S16 in Table 1). Figure 1 shows a sketch of the structure and the conduction band profile of one of the QWs, including the location of the first and second quantum-confined electron levels and their associated squared wavefunctions. Note that the potential profile of the MQW displays the triangular pattern characteristic of GaN/AlN heterostructures, due to the internal electric field associated to the difference in spontaneous and piezoelectric polarization between the wells and the barriers. The electric field results in a redshift of the band-to-band transitions (quantum confined Stark effect) and a blueshift of the ISB transitions with respect to square MQWs [24]. Additionally, the fact that the gravity centers of the wavefunctions associated to the first hole level and the first and second electron confined levels are shifted along the growth axis leads to a decrease of the dipole matrix elements associated to both of the band-to-band and ISB transitions. Growth along nonpolar crystallographic orientations, such as the <1-100> *m* axis, inhibits the appearance of internal electric field, with the resulting enhancement of the dipole matrix elements. However, the lattice mismatch in the nonpolar planes is larger, which leads to a high density of dislocations, cracks and stacking faults when the average Al concentration in the structure exceeds ≈ 10% [42]. Therefore, in architectures for ISB optoelectronics, the use of nonpolar crystallographic orientations is restricted to applications in the FIR.

After growth, the periodicity of the structure was confirmed by measuring ω-2θ scans of the (0002) x-ray reflection. Typical diffractograms for Ge-doped and Si-doped structures are presented in Fig. 1(c), together with a simulation performed with the Rigaku SmartLab Studio software. The simulation provides a good reproduction of the experimental result assuming that the MQW presents the in-plane lattice parameters of an AlGaN layer with the average Al concentration of the MQW, i.e. the MQW structure is fully relaxed. This is expected to be the case in view of the thickness of the MQW, following the study in ref. [43]. The values of the MQW period extracted from the intersatellite angular distance in the diffraction pattern are summarized in Table 1. Also, the full width at half maximum (FWHM) of the rocking curves was measured for the AlN substrate and the MQW zero-order reflection [MQW 0 peak in Fig. 1(c)], obtaining an average FWHM of $0.19 \pm 0.01°$ for Si-doped MQWs and $0.16 \pm 0.01°$ for Ge-doped MQWs, to be compared with $0.057 \pm 0.006°$ for the AlN-on-sapphire template. The broadening of the MQW rocking curve in comparison with the substrate is due to strain relaxation. On the other hand, Ge-doped MQWs systematically present ≈18% narrower rocking curves than their Si-doped counterparts for all the doping levels. This correlation indicates that, for similar substrate quality, Ge-doping results in structures with better mosaicity than when using Si-doping.

To assess the band-to-band properties of the MQWs, the PL spectra of all the samples were measured at 5 K. The results in terms of peak emission energy are shown in Table 1, and the spectra are displayed in Figs. 2(a) and (b). Both for Ge- and Si-doping, the spectra present a multipeak structure resulting from monolayer fluctuations of the well thickness [8]. Increasing the dopant density induces a blueshift of the PL energy, which is assigned to the screening of the internal electric field by the free carriers [44], as well as a broadening of the emission peak, which is due to the Burstein-Moss effect [45]. The total shift in emission energy and the energy broadening are similar for both dopants. Calculated values of the band-to-band transition energy obtained using the nextnano3 software are indicated by black triangles in the figure. These calculations, which take into account the sample period and the screening of the internal electric fields by free carriers, provide a reasonable fit to the experimental results. The agreement confirms the incorporation and activation of both Si and Ge in the QWs.

The vertical homogeneity of the MQW stack has been studied by low-temperature (10 K) CL, recording a spectral line-scan on a cross-section of sample S16 cleaved along the growth direction. Figure 3 displays the obtained spectral map, along with a sketch of the sample. Following the growth axis, the MQW emission broadens, redshifts and gains intensity, which is explained by the gradual relaxation of misfit strain induced by the AlN substrate, which can extend over 10-20 MQW periods [43]. Theoretical calculations predict a redshift of 11 nm when the superlattice evolves from being fully strained on AlN to fully relaxed.

To assess the ISB properties of the samples, the absorption in the SWIR range was measured at room temperature by FTIR spectroscopy. Figures 2(c) and (d) show the normalized absorption spectra for TM-polarized light. As explained in the PL case, the multipeak structure of the spectra is due to monolayer thickness fluctuations in the wells [8]. The ISB absorption peak energies for the various samples are presented in Table 1, together with the magnitude of the absorption per pass. The value of absorption should be taken cautiously, since the accumulated error associated to the calculation of the waveguide length and incident angle can reach ±20%. However, comparing S11-12 and S13-14, we can conclude the magnitude of the absorption scales linearly with the doping density, as expected, and a saturation is observed for S15-16, which is explained by the Fermi level approaching the excited state $e_2$ in these heavily doped structures. As expected, the absorption resonances are blueshifted when increasing the doping concentration, which is due to many-body effects, namely exchange interaction and depolarization shift. The magnitude of these effects has been quantified as described in ref. [46], giving the theoretical transition energies indicated by black triangles in Fig. 2. In general, the magnitude of the spectral shift agrees well with the theoretical expectations for both Si and Ge. Regarding the absorption linewidth, we observe that Si-doping leads to stronger broadening of the absorption peak, as summarized in Fig. 2(e). Thus, in this case, broadening is not dominated by the scattering by ionized impurities or electron-electron interaction, but rather by interface roughness.

**B. MIR absorption in *c*-plane GaN/AlGaN MQWs**

To extend the study to the MIR range, the AlN barriers were replaced by AlGaN, and the width of the QWs was increased. Therefore, this study focuses on structures consisting of 30 periods of GaN/Al$_{0.33}$Ga$_{0.67}$N (4 nm/3 nm) MQWs, designed to show ISB absorption at 240 meV (5.2 µm) at low doping levels (samples S21 to S26 in Table 1). A 2-nm-thick region at the center of the GaN QWs was homogeneously doped with Si or Ge, to reach the surface dopant densities described in Table 1. Figure 4 shows a sketch of the structure as well as the conduction band diagram of one of the QWs in the middle of the stack. Typical HR-XRD ω–2θ scans of the (0002) reflection for these samples are shown in Fig. 4(c). The diffractograms are complex due to the presence of an additional multiple-heterostructure close to the silicon/GaN interface with a period of ≈ 20 nm. The purpose of this structure is to maintain the GaN under compressive strain during the cool down process after growth, thus preventing crack propagation. The simulation presented in the figure does not take into account the buffer layer, and assumes that the MQW presents the in-plane lattice parameters of an AlGaN layer with the average Al concentration of the MQW, as it is expected [47]. By comparison of the experimental diffractograms with the simulation, it is possible to identify the satellites corresponding to the PAMBE-grown GaN/AlN MQW structure and determine its period. The results are summarized in Table 1. Additionally, the structural quality was evaluated by measuring the FWHM of the rocking curves for the GaN template and the MQW zero-order reflection. The resulting data, also listed in Table 1, yields average values of $0.189 \pm 0.005°$ and $0.187 \pm 0.005°$ for Si-doped and Ge-doped MQWs, respectively, to be compared with $0.208 \pm 0.005°$ for the GaN-on-Si(111) templates. We can hence conclude that the mosaicity of the samples shows no clear dependence on either the dopant nature or the doping density, and is rather determined by the substrate.

Concerning the band-to-band properties of the MQWs, the low temperature (5 K) PL spectra from samples with different doping levels of Ge and Si are presented in Figs. 5(a) and (b), respectively. Increasing the Ge-doping density results in a slight blueshift of the PL energy, as well as an asymmetric widening of the PL peak towards the low-energy side. The magnitude of the spectral shift, associated with the screening of the internal electric field, is significantly smaller than in the samples S11-S16 described in the previous section, in spite of the fact that the QWs are larger (4 nm in S21-S26 to be compared with 2 nm in S11-S16). This reduced shift is mainly due to the reduction of the internal electric fields when using $Al_{0.33}Ga_{0.67}N$ barriers instead of AlN. Looking at the PL spectra in Figs. 5(a-b), the emission from S25 appears surprisingly narrow for a sample with such a high doping level. A closer look at the spectrum reveals that the main emission consists of two peaks: a dominant "narrow" line at 3.46 eV and a broader emission at higher energies, around 3.51 eV, with similar characteristics (in logarithmic scale, not shown) than the emission from S26. Comparing the two samples, the emission at 3.46 eV in S25 is probably due to localized emission at structural inhomogeneities, but we did not manage to identify such defects in transmission electron microscopy images.

To assess the homogeneity of the MQW stack along the growth axis, a spectral line-scan of the low-temperature (10 K) CL was recorded on a cross-section of the sample with the highest Ge doping level (S26) cleaved along the growth direction (Fig. 6). In agreement with the PL experiments in Fig. 5(a), the emission from the MQW, peaking at 355 nm (3.49 eV), is strongly asymmetric. Its intensity increases along the growth axis, but the spectral shift is reduced in comparison with the GaN/AlN MQWs (see Fig. 3), which is consistent with the smaller lattice mismatch with the substrate. The luminescence from the AlGaN capping layer is visible as an extra emission peak centered around 327 nm (3.79 eV). There is no detectable emission from the AlGaN barriers, which points to a good charge transfer from the barriers to the wells.

The ISB properties of the MQWs were measured by FTIR at room temperature. As summarized in Table 1 and illustrated in Figs. 5(c) and (d), all samples present a TM-polarized absorption peak, which blueshifts for increasing doping levels. As in the SWIR set, the absorption scales sublinearly with the doping density, which is justified by the high doping levels that bring the Fermi level close to $e_2$, or even beyond. The spectral location of the absorption resonance fits well with theoretical calculations of the ISB transition taking the exchange interaction and depolarization shift into account (indicated with black triangles in the figures). When increasing the dopant concentration, the absorption peak broadens significantly. As discussed for GaN/AlN MQWs, the fact that the broadening is more pronounced for the Si-doped samples [see Fig. 5(e)] points to an increase of the interface roughness or the density of structural defects, although no clear evidence of these facts were identified in high-resolution transmission electron microscopy measurements.

## C. FIR absorption in *m*-plane GaN/AlGaN MQWs

The design of *c*-plane GaN/AlGaN MQWs displaying ISB transitions in the FIR is hindered by the polarization-induced internal electric field, which leads to an increased carrier confinement in the QWs. Thus, nonpolar crystallographic orientations are the most promising choice for the development of a GaN-based ISB technology in the FIR range, and we have decided to focus on the *m* plane to study the effect of Ge-doping. The 40-period *m*-plane GaN/Al$_{0.075}$Ga$_{0.925}$N (10 nm/18.5 nm) MQWs presented here are reproduced from a previously-reported Si-doped series [29], designed so that, for low doping densities, two electron levels are clearly confined in the QWs, e$_1$ and e$_2$, separated by ≈ 30 meV (41 µm, 7.3 THz), and the third electronic level, e$_3$, is located almost at the conduction band edge of the barriers. In this case, the GaN wells were homogeneously doped with Ge, at concentrations increasing from $5 \times 10^{11}$ cm$^{-2}$ to $5 \times 10^{12}$ cm$^{-2}$, as described in Table 1. Figures 7(a) and (b) show the schematic description of the samples and the conduction band diagram of one of the QWs in the middle of the stack, indicating the energy of e$_1$, e$_2$, and e$_3$ and the associated squared wavefunctions. Figure 7(c) shows the HR-XRD ω–2θ scan of the (3-300) reflection of sample S33, containing the most heavily doped MQWs. The thicknesses of the MQW periods for all the samples of the series extracted from the intersatellite distance of such ω–2θ scans are listed in Table 1, together with the FWHM of the rocking curves for the bulk GaN substrate and the MQW zero-order reflection (0.011 ± 0.001° and 0.013 ± 0.001°, respectively). The values of the FWHM are very similar for the substrate and for the MQW, in agreement with the Si-doped series [29]. In view of these results, we conclude that neither the dopant nature nor the doping density modify the MQW mosaicity, which is rather determined by the substrate.

As an additional evaluation of the structural quality, we recorded AFM images of the samples, as illustrated in Fig. 8(a). The surfaces are characterized by atomic terraces, with a root-mean-square (rms) roughness < 0.3 nm in 5×5 µm$^2$ scans. Sample S32 was also characterized by HAADF-STEM, as shown in Figs. 8(b) and (c), where layers with dark and bright contrast correspond to the AlGaN barriers and GaN QWs, respectively. Cross-section images of the stack do not show any extended defects (neither dislocations nor stacking faults) over an in-plane length of ≈2 μm. Figure 8(c) shows two QWs in the middle of the sample viewed along the <0001> zone axis. In the barriers, slight alloy inhomogeneities appear in the form of darker lines parallel to the QW interface, as observed in similar Si-doped structures [25].

The low-temperature (5 K) PL of the nonpolar MQWs is displayed in Fig. 9(a). The spectra consist of a main emission peak at 3.48 eV originating from the MQW and a weaker emission about 100 meV higher in energy that is assigned to the $Al_{0.075}Ga_{0.925}N$ cap layer. The effect of the doping density on the emission energy and broadening is negligible in the range under study. This is due to the fact that, first, in nonpolar QWs there is no electric field to be screened by carriers, and second, the surface dopant densities described in Table 1 should not have a large effect on the band-to-band transition (the shift due to band filling is < 60 meV, and it is partially compensated by bandgap renormalization [48]). The analysis of the CL spectral line-scan on the cross-section of the sample, presented in Fig. 10, confirms that the $Al_{0.075}Ga_{0.925}N$ emission originates from the cap layer and not from the barriers, where the generated carriers are fully transferred to the QWs. The emission from the QWs is homogeneous in linewidth and intensity all along the structure, as it was also the case for Si-doped QWs with the same structure and similar dopant density [29].

The ISB properties of the structures were assessed by FTIR, and compared to the data obtained in ref. [29] from Si-doped MQWs. In Fig. 9(b), we show the normalized absorption spectra for TM-polarized light from the three Ge-doped samples (peak energies are summarized in Table 1), and the normalized broadening of both the Si-doped and the Ge-doped series is compared in Fig. 9(c). The absorption spectra do not vary significantly with increasing Ge doping levels. This behavior is unexpected since in sample S32 the Fermi level should already reach the second electron subband, as confirmed by the saturation of the ISB absorption per pas in Table 1, which should also lead to enhanced scattering. For equivalent doping densities, the Ge-doped structures show ISB absorption peaks with slightly smaller broadening than that of the Si-doped structures. When increasing the Ge-doping density, we observe a slight widening of the ISB absorption peak and no shift of the ISB energy.

## 4. Discussion and conclusions

In summary, the structural and optical properties of Ge-doped GaN/AlGaN MQWs designed to display ISB absorption in the SWIR, MIR and FIR ranges have been characterized and compared with similar structures doped with Si. For this purpose, we have grown 25-period *c*-plane GaN/AlN (1.8 nm/3 nm) MQWs designed to show ISB absorption at 1.70 μm (0.729 eV), 30-period *c*-plane GaN/Al$_{0.33}$Ga$_{0.67}$N (4 nm/3 nm) MQWs designed to display ISB absorption at 5.2 μm (240 meV), and 40-period *m*-plane GaN/Al$_{0.075}$Ga$_{0.925}$N (10 nm/18.5 nm) MQWs designed to show ISB absorption at 41 μm (30 meV).

The results presented above confirm the feasibility of using Ge instead of Si in ISB optoelectronic devices consisting of GaN/AlN or GaN/AlGaN heterostructures. However, we can debate on the relevance of using one or the other dopant. For this comparison, it must be kept in mind that Si and Ge have approximately the same activation energy in GaN [49], and none of them perturbs the growth kinetics of GaN when using PAMBE [35,40]. However, Si is known to introduce a strong local deformation of the GaN lattice [50] as well as an enhancement of the interface roughness in GaN/AlGaN QWs [31]. Looking at our results, Si-doped and Ge-doped GaN/AlGaN MQWs are structurally similar both in *c*-plane and *m*-plane crystallographic orientations, with the morphology and mosaicity being determined rather by the substrate and not by the dopant nature or density (in the range under consideration). Only in strongly lattice-mismatched MQWs (GaN/AlN), which exhibits a clear structural degradation with respect to the substrate, Ge results in a systematic structural improvement.

If we turn to the optical properties, the evolution of the band-to-band behavior as a function of doping reflects the screening of the internal electric field by free carriers, which is independent of the nature of the dopant. The ISB processes, on the other hand, are more sensitive to parameters like the roughness of the heterointerfaces. Here, results for low doping levels are comparable for MQWs doped with Si or Ge. However, for high doping levels, there is a systematic improvement when using Ge as a dopant, which manifests in narrower absorption bands independent of the spectral region, and thus for different QW size, barrier composition and crystallographic orientations.

## Acknowledgements

The authors acknowledge technical support from Y. Curé and Y. Genuist. Thanks are due to N. Mollard for sample preparation by focused ion beam at the NanoCharacterization Platform (PFNC) in CEA-Minatec Grenoble. The free-standing semi-insulating *m*-GaN substrates were kindly supplied by Suzhou Nanowin Science and Technology Co. Ltd (Nanowin). This work is supported by the EU ERC-StG "TeraGaN" (#278428) project. A.A. acknowledges financial support by the French National Research Agency via the GaNEX program (ANR-11-LABX-0014).

## Tables

**Table 1.** Structural and optical characteristics of the GaN/AlGaN MQWs under study: dopant nature; surface dopant density; thickness of the MQW period extracted from HR-XRD; FWHM of the ω-scan of the (0002) and (3̄300) x-ray reflections for the *c*-plane and *m*-plane samples, respectively, of the MQWs and of the AlN or GaN substrate; peak energy of the MQW PL emission at $T = 5$ K; peak energy of the ISB absorption; FWHM of the ISB absorption peak; ISB absorption per pass.

| Sample | Dopant nature | Doping concentration ($\times 10^{12}$ cm$^{-2}$) | Period thickness (nm) | ω-scan FWHM MQW (°) | ω-scan FWHM substrate (°) | PL peak position (eV) | ISB absorption energy (meV) | ISB absorption FWHM (meV) | ISB absorption per pass (%) |
|---|---|---|---|---|---|---|---|---|---|
| S11 | Si | 6 | 4.8±0.1 | 0.199 | 0.056 | 3.37 | 710 | 85 | 2.1 |
| S12 | Ge | 6 | 4.8±0.1 | 0.167 | 0.046 | 3.35 | 720 | 94 | 1.8 |
| S13 | Si | 20 | 4.8±0.1 | 0.180 | 0.061 | 3.45 | 830 | 108 | 8.9 |
| S14 | Ge | 20 | 4.8±0.1 | 0.154 | 0.058 | 3.46 | 760 | 90 | 6.6 |
| S15 | Si | 60 | 4.4±0.1 | 0.176 | 0.057 | 3.51 | 860 | 189 | 9.5 |
| S16 | Ge | 60 | 4.3±0.1 | 0.149 | 0.063 | 3.63 | 820 | 110 | 8.6 |
| S21 | Si | 2 | 7.0±0.1 | 0.190 | 0.212 | 3.41 | 249 | 45 | 4.7 |
| S22 | Ge | 2 | 7.0±0.1 | 0.188 | 0.209 | 3.41 | 272 | 59 | 4.1 |
| S23 | Si | 10 | 7.0±0.1 | 0.187 | 0.210 | 3.42 | 286 | 80 | 5.8 |
| S24 | Ge | 10 | 7.0±0.1 | 0.185 | 0.210 | 3.41 | 276 | 77 | 7.4 |
| S25 | Si | 20 | 6.9±0.1 | 0.189 | 0.206 | 3.43 | 280 | 140 | 15 |
| S26 | Ge | 20 | 6.9±0.1 | 0.187 | 0.204 | 3.49 | 273 | 90 | 16 |
| S31 | Ge | 0.5 | 27.5±0.1 | 0.014 | 0.012 | 3.48 | 25 | 12 | 19 |
| S32 | Ge | 2 | 28.2±0.1 | 0.013 | 0.010 | 3.48 | 24 | 13 | 19 |
| S33 | Ge | 5 | 29.3±0.1 | 0.013 | 0.011 | 3.48 | 24 | 15 | 13 |

**Figure captions**

**Figure 1.** (a) Schematic of the GaN/AlN MQW structures. (b) Band diagram of one of the QWs and location of the first ($e_1$) and second ($e_2$) electron levels obtained using a self-consistent 8-band k.p Schrödinger-Poisson solver. (c) HR-XRD ω-2θ scan of the (0002) reflection of Si- and Ge-doped heterostructures (S15 and S16, respectively). Experimental data are compared to a simulation which assumes that the MQWs present the in-plane lattice parameters of an AlGaN layer with the average Al concentration of the MQW.

**Figure 2.** Low temperature ($T$ = 5 K) PL intensity of the (a) Ge-doped and (b) Si-doped GaN/AlN MQWs. The spectra are normalized by their maximum value and vertically shifted for clarity. Black triangles mark the transition energies calculated using a self-consistent 8-band k.p Schrödinger-Poisson solver that takes into account the screening of the internal electric field by free carriers. Normalized TM-polarized ISB absorption of the (c) Ge-doped and (d) Si-doped MQWs measured by FTIR spectroscopy. The spectra are normalized by their maximum value and vertically shifted for clarity. Black triangles mark the transition energies calculated using a self-consistent 8-band k.p Schrödinger-Poisson solver and corrected to account for the exchange interaction and depolarization shift. (e) Normalized broadening (ΔE/E) of the ISB absorption peak as a function of the doping density ($N_S$).

**Figure 3.** CL spectral line-scan on the cross-section of sample S16 measured at $T$ = 10 K. The intensity is color-coded on a logarithmic scale. The sketch on the right side shows the corresponding sequence of layers.

**Figure 4.** (a) Schematic of the GaN/Al$_{0.33}$Ga$_{0.67}$N MQW structures. (b) Band diagram of one of the QWs and location of the first ($e_1$) and second ($e_2$) electron levels obtained using a self-consistent 8-band k.p Schrödinger-Poisson solver. (c) HR-XRD ω-2θ scan of the (0002)

reflection of Si- and Ge-doped heterostructures (S25 and S26, respectively). Experimental data are compared to a simulation which assumes that the MQWs present the in-plane lattice parameters of an AlGaN layer with the average Al concentration of the MQW. The simulation does not take into account the buffer layer of the GaN-on-Si(111) template. This buffer contains a periodic GaN/AlGaN heterostructure (period ≈ 20 nm).

**Figure 5.** Normalized PL intensity of the (a) Ge-doped and (b) Si-doped GaN/Al$_{0.33}$Ga$_{0.67}$N MQWs measured at $T = 5$ K. The spectra are normalized by their maximum value and vertically shifted for clarity. Black triangles mark the transition energies calculated using a self-consistent 8-band k.p Schrödinger-Poisson solver that takes into account the screening of the internal electric field by free carriers. Normalized TM-polarized ISB absorption of the (c) Ge-doped and (d) Si-doped MQWs measured by FTIR spectroscopy. The spectra are normalized by their maximum value and vertically shifted for clarity. Black triangles mark the transition energies calculated using a self-consistent 8-band k.p Schrödinger-Poisson solver and corrected to account for the exchange interaction and depolarization shift. (e) Normalized broadening ($\Delta E/E$) of the ISB absorption peak as a function of the doping density ($N_S$).

**Figure 6.** CL spectral line-scan on the cross-section of sample S26 measured at $T = 10$ K. The intensity is color-coded on a logarithmic scale. The sketch on the right side shows the corresponding sequence of layers.

**Figure 7.** (a) Schematic of the *m*-plane GaN/Al$_{0.075}$Ga$_{0.925}$N MQW structures. (b) Band diagram of one of the QWs and location of the first ($e_1$), second ($e_2$) and third ($e_3$) electron levels obtained using a self-consistent 8-band k.p Schrödinger-Poisson solver. (c) HR-XRD ω-2θ scan of the (3-300) reflection of S33. Experimental data are compared to a simulation which assumes that the MQWs are fully strained on GaN.

**Figure 8.** (a) AFM image (5×5 μm$^2$) of sample S31 showing an rms surface roughness of 0.28 nm. (b)-(c) Cross-section HAADF-STEM images of sample S32 viewed along <0001>. Layers with dark and bright contrast correspond to the AlGaN barriers and GaN QWs, respectively.

**Figure 9.** (a) Low temperature ($T$ = 5 K) PL intensity of the Ge-doped *m*-plane MQWs. The spectra are normalized by their maximum value and vertically shifted for clarity. (b) Normalized absorption for TM-polarized light measured at 5 K by FTIR spectroscopy. (c) Normalized broadening (ΔE/E) of the ISB absorption peak as a function of the doping density ($N_S$). The triangular data points correspond to the Ge-doped MQWs under study, whereas the square data points correspond to similar Si-doped MQWs from ref. [29].

**Figure 10.** CL spectral line-scan on the cross-section of sample S33 measured at $T$ = 10 K. The intensity is color-coded on a logarithmic scale. The sketch on the right side shows the corresponding sequence of layers.

**Figure 1**

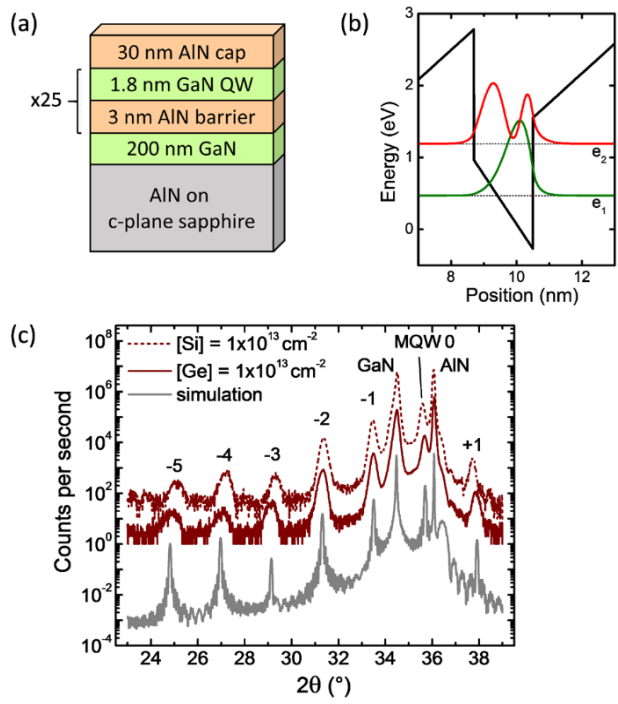

**Figure 2**

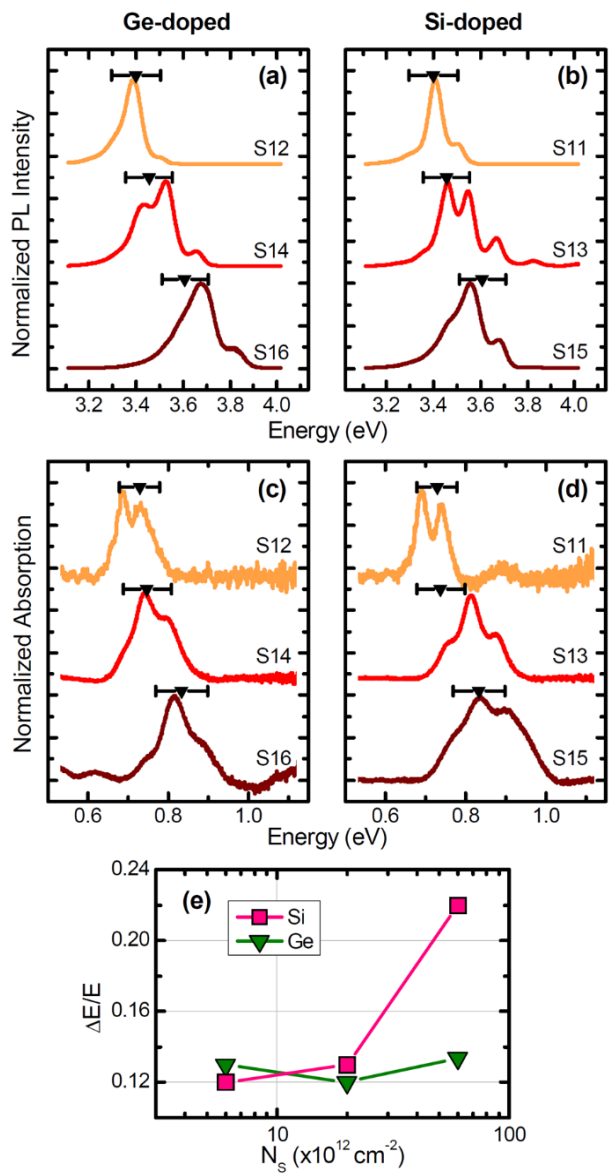

**Figure 3**

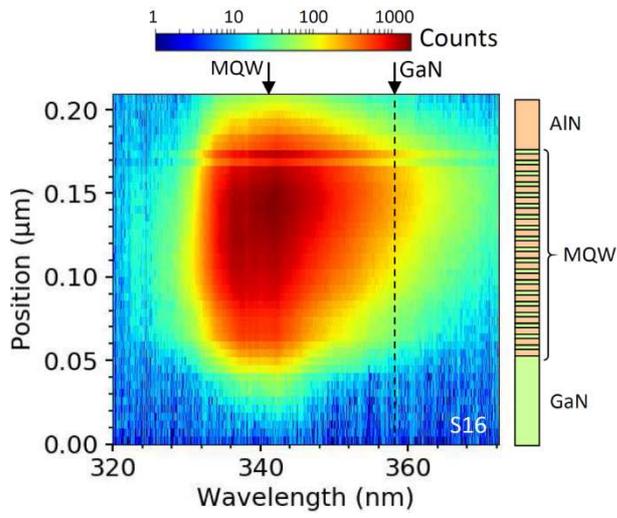

**Figure 4**

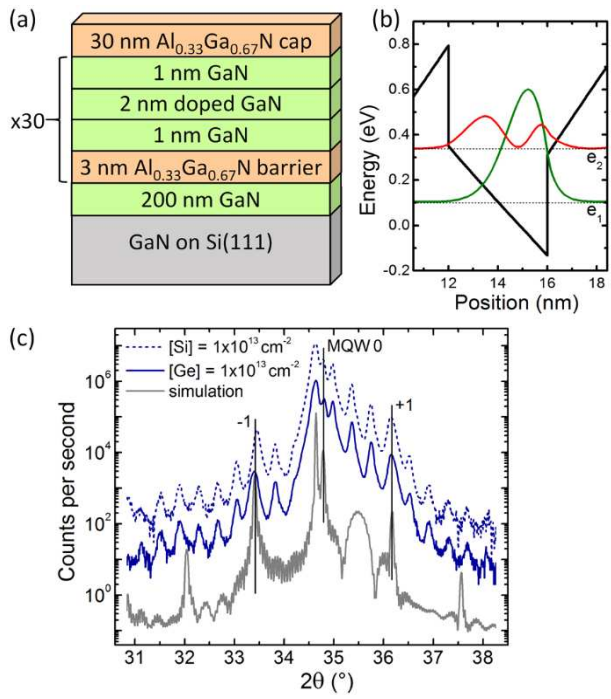

**Figure 5**

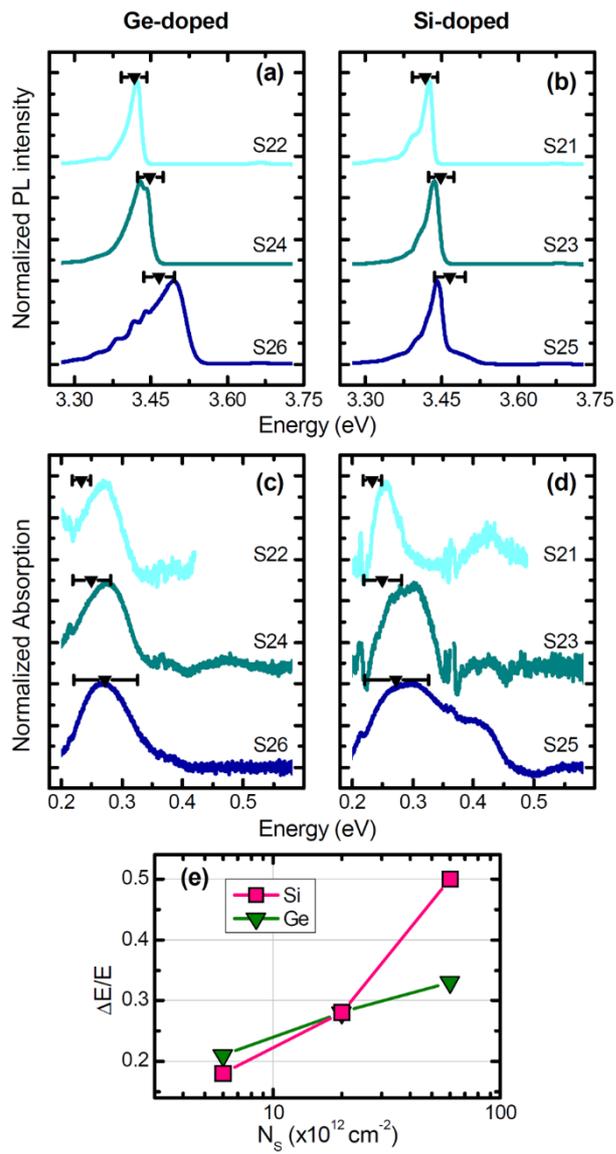

**Figure 6**

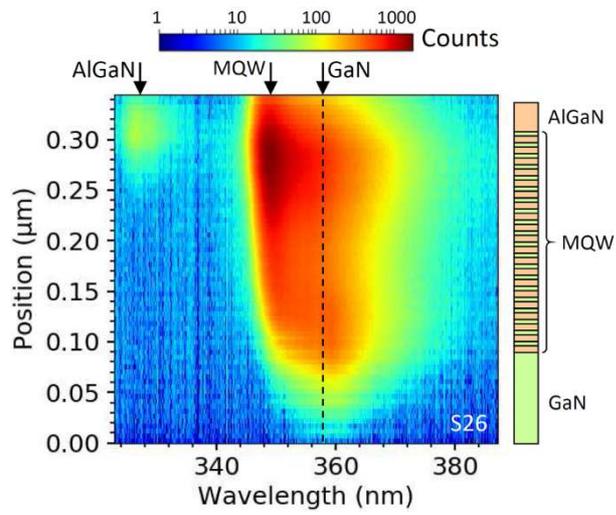

**Figure 7**

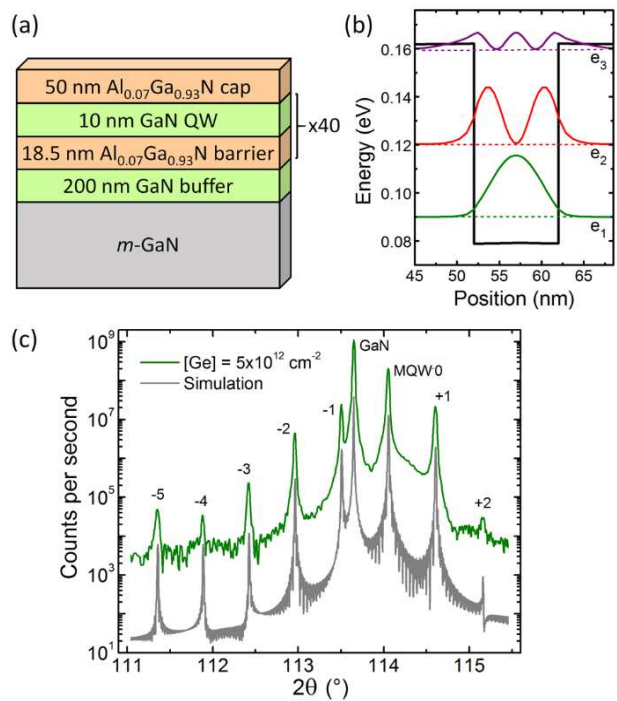

**Figure 8**

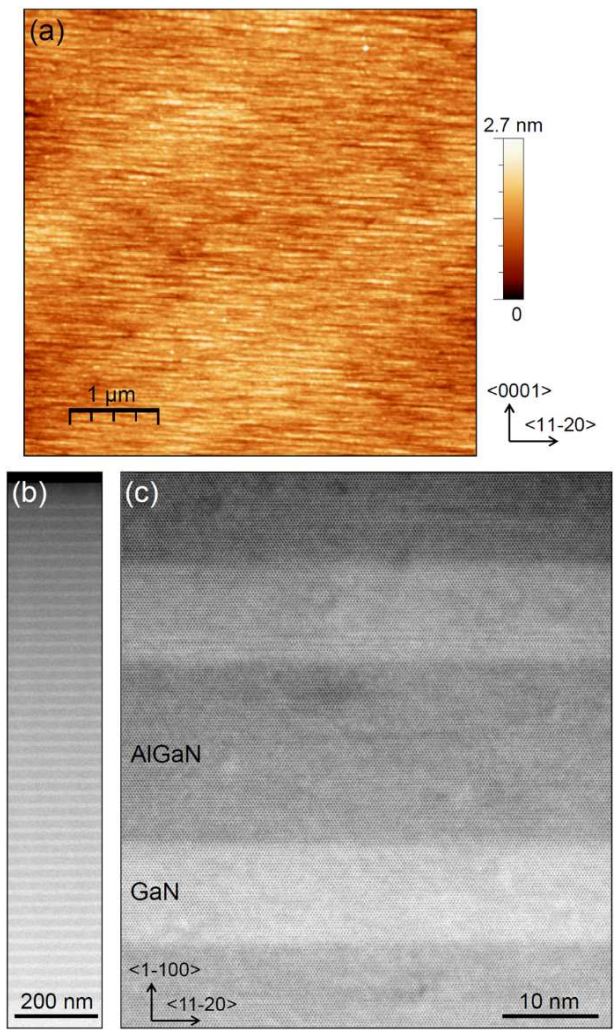

**Figure 9**

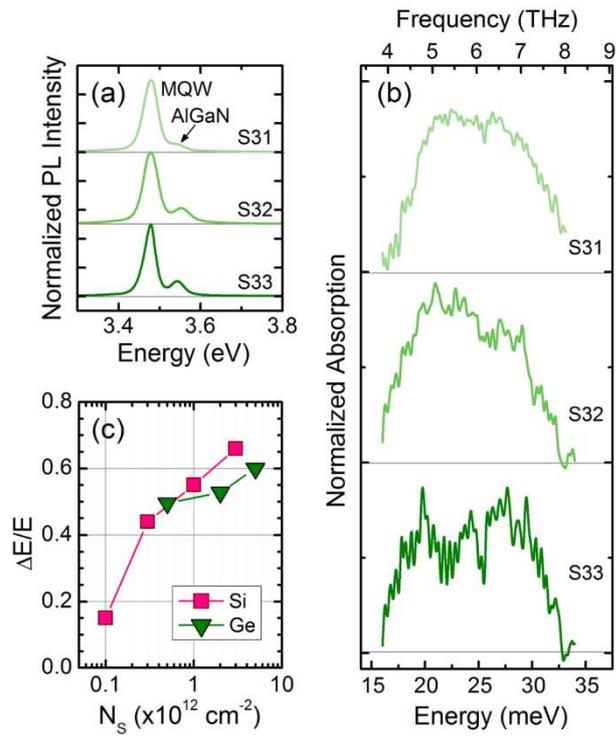

**Figure 10**

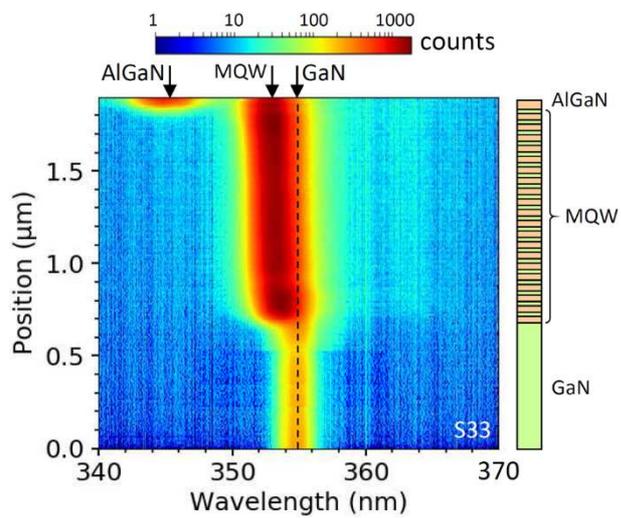